\documentstyle[12pt,epsf]{article}

\begin{document}

\thispagestyle{empty}

\font\fortssbx=cmssbx10 scaled \magstep2
\hbox to \hsize{
\hskip.5in \raise.1in\hbox{\fortssbx University of Wisconsin - Madison}
\hfill$\vcenter{\hbox{\bf MADPH-96-970}
                \hbox{November 1996}}$ }

\vspace{.5in}

\begin{center}{\large\bf
Universal Behavior in Excited Heavy-Light\\ and Light-Light Mesons}\\[5mm]
M. G. Olsson\\
\it Physics Department, University of Wisconsin, Madison, WI 53706
\end{center}

\vspace{1in}

\begin{abstract}
A common pattern of large orbital and radial excitations in heavy-light and light-light mesons is demonstrated. Within a general potential model the Regge slopes of the light degrees of freedom for these mesons are shown to be in the ratio of two. The possibility of ``tower" degeneracy occurs only with pure scalar confinement.
\end{abstract}

\newpage

\section{Introduction}

Although only partially explored, the high orbital and radial excitations of mesons promise many exciting possibilities. Highly excited mesons are analogous to the well known ``Rydberg atom" which for decades has been a useful tool to investigate atomic dynamics. In a similar way, highly excited meson states sample the interaction at large distance and can be treated semi-classically.

It is often remarked that the quark-antiquark interaction ``saturates" at large distance to reflect the possibility of string breaking. That this is not the case is clearly seen by the existence of linear Regge trajectories. For example, the ``rho" trajectory has states at least out to $J=5$\cite{pdg} with no obvious departure from linearity. As we emphasize again here, linear Regge trajectories result from light quark kinematics and linear confinement. In this case the orbital radius is about 7~GeV$^{-1}$ (for the $J=5$ state) and the confinement energy is 1.4~GeV, well above that needed for string breaking.
A more correct concept of large distance confinement is that the interquark potential continues to rise linearly with coupled hadron channels implying widths to the meson states.

The theoretical description of dominantly orbital states (DOS) has some attractive simplifying features. The leading Regge trajectory is ``classical" and radially excited states can be treated semi-classically. Furthermore, these DOS do not penetrate to small distances so that only the confining interaction needs to be considered.

In this paper the DOS regime is explored with light quark kinematics and an arbitrary superposition of time component vector and scalar confinement. A feature of this work is the comparison between the global DOS spectra in the heavy-light limit compared to mesons containing two light quarks. We find that in terms of the angular momentum $J$ and the radial quantum number $n$, the mass of the light degrees of freedom in both cases is given by
\begin{equation}
\alpha' M^2 = J + (2n+1)C \,,  \label{lightmass}
\end{equation}
where $\alpha'$ is the Regge slope. Both $\alpha'$ and $C$ depend on the nature of the confinement (i.e., the fraction of scalar confinement). The energy $M$ is the meson mass in the light-light case but one must add the heavy quark mass in the heavy-light case.

The universal aspect of the spectroscopy (\ref{lightmass}) is that for a given confinement choice,
\begin{equation}
\begin{array}{l}
\alpha'(\mbox{heavy-light}) = 2\alpha'(\mbox{light-light}) \,,\\
C(\mbox{heavy-light}) = C(\mbox{light-light}) \,.
\end{array}
\end{equation}
Thus, by a factor of two rescaling of $M^2$ the light-light and heavy-light spectroscopies become identical, at least in the DOS regime. In Section~2 we establish our semi-classical formalism in the heavy-light limit. Detailed predictions for a mixed vector and scalar linear confinement are worked out in Section~3. The two light quark meson case is considered in Section~4. We summarize our results in Section~5. Some results for light-light mesons have been presented earlier\cite{goebel}.

\section{Orbital Dominant Heavy-Light Mesons}

A classical quark of mass $m$ moving with velocity {\bf v} in an effective scalar potential $S(r)$ and a time-component vector potential $V(r)$ with a static source at the origin $(r=0)$ is governed by the Lagrangian,
\begin{equation}
L = \bigl(m+S(r)\bigr) \sqrt{1-v^2} - V(r) \,,
\end{equation}
where $v^2=\dot r^2 + \omega^2r^2$. By the usual methods we obtain the canonical momenta
\begin{eqnarray}
p_r &=& {\partial L\over\partial\dot r} = (m+S)\gamma\dot r\,,\\
J &=& {\partial L\over\partial \omega} = (m+S)\gamma\omega r^2 \,,
\end{eqnarray}
where $\gamma=(1-v^2)^{-1/2}$ and the Hamiltonian is then
\begin{equation}
H = \dot r p_r + \omega J - L = (m+S)\gamma + V \,. \label{ham}
\end{equation}
By observing that 
\begin{equation}
p^2 = p_r^2 + {J^2\over r^2} = (m+S)^2 \gamma^2 v^2 = (m+S)^2 (\gamma^2-1)
\end{equation}
the velocities can be eliminated in (\ref{ham}) to obtain
\begin{equation}
H = \sqrt{\bigl(m+S(r)\bigr)^2 + p_r^2 + {J^2\over r^2}} \; + V(r) \,. \label{class-ham}
\end{equation}

\bigskip\noindent
\underline{\bf Circular orbits}

For large orbital angular momentum $J$ the lowest energy orbit will be circular. From (\ref{ham}) we define
\begin{equation}
H(r=r_0,\ p_r=0,\ J) \equiv M_0 = (m+S_0) \gamma_0 + V_0 \,.
\end{equation}
The circular orbit condition (for $p_r=0$) is
\begin{equation}
\left. {\partial L\over\partial r}\right|_{r=r_0} = 0 = (m+S_0) \gamma_0 \omega_0^2 r_0 - V_0' - {S_0'\over\gamma_0} \,, \label{circ-orbit}
\end{equation}
where $V'={dV\over dr}$ etc. Using $\gamma_0^2 \omega_0^2 r_0^2 = \gamma_0^2-1$ we find
\begin{eqnarray}
M_0 &=& {1\over2} \left.\left( V' + {V\over r} + \sqrt{(V')^2 + 4S (S+rS')/r^2} \right)\right|_{r=r_0} r_0 \,, \label{M_0} \\[4mm]
J &=& \left.\left[{1\over2}V' \left(V'+\sqrt{(V')^2 + 4S(S+rS')/r^2}\right) + {SS'\over r}\right]\right|_{r=r_0} r_0^2 \,.  \label{J}
\end{eqnarray}

\bigskip\noindent
\underline{\bf Radial oscillations}

For a given (non-zero) angular momentum $J$, Eq.~(\ref{J}) can be used to solve for the classical circular orbit radius. For linear confinement, as we will see in the next section, this becomes trivial. Radial oscillations will in general occur and the harmonic oscillator approximation is equivalent to a semi-classical JWKB treatment for large $J$. To obtain the oscillator expansion we square the classical Hamiltonian (\ref{class-ham}),
\begin{equation}
(M-V)^2 = (m+S)^2 + {J^2\over r^2} + p_r^2 \,, \label{class-ham^2}
\end{equation}
where we replace $H$ by its eigenvalue $M$. We subtract from this the circular version of (\ref{class-ham^2})
\begin{equation}
(M_0-V_0)^2 = (m+S_0)^2 + {J^2\over r^2_0}
\end{equation}
and expand to first order in $M-M_0$ and second order in $\Delta r = r-r_0$,
\begin{eqnarray}
p_r^2 &=& 2(M_0-V_0) (M-M_0) + 2 \left[ {J^2\over r_0^3} - (m+S_0)S'_0 + (M_0-V_0)V'_0 \right] \Delta r \nonumber\\
&&\quad {}+ \left[ V'^2_0 - (M_0-V_0)V_0'' - S'^2_0 - (m+S_0)S_0'' - 3{J^2\over r_0^4} \right] (\Delta r)^2 \,.
\end{eqnarray}
As expected, the coefficient of $\Delta r$ is equivalent to the circular orbit condition (\ref{circ-orbit}) and hence vanishes. The radial motion equation is simple harmonic with classical frequency
\begin{equation}
\omega = {\left[ -(V_0')^2 + (M_0-V_0)V_0'' + (S_0')^2 + (m+S_0)S_0'' + 3J^2/ r_0^4 \right]^{1/2} \over M_0-V_0} \label{omega}
\end{equation}
and hence the quantized eigenvalue is
\begin{equation}
M = M_0 + \omega\left(n+{1\over2}\right)\,, \qquad n = 0,1,2,\dots
\end{equation}
For DOS states $M_0\gg\omega$ so squaring (and dropping the $\omega^2$ term) gives
\begin{equation}
M^2 = M_0^2 + M_0 \omega (2n+1) \,. \label{M^2}
\end{equation}

\section{Heavy-Light Mesons with General Linear Confinement}

Our general confinement model here consists of a combination of scalar and time-component vector confinement, both assumed to be of the linear type
\begin{equation}
\begin{array}{l}
S(r) = far\,,\\
V(r) = (1-f)ar\,,
\end{array} \label{genlinconf}
\end{equation}
where $f$ is a dimensionless constant and $a$ is the ``string tension". With the assumption of linear confinement and setting the quark mass to zero, the expressions for $M_0$ and $J$ in (\ref{M_0}) and (\ref{J}) simplify to
\begin{eqnarray}
M_0 &=& (\lambda+1-f)ar_0\,, \label{M_0:simple} \\
\lambda &=& {1\over2}\left(1-f+\sqrt{(1-f)^2+8f^2}\right)\,,\\
J &=& \nu a r_0^2\,, \label{J:simple}\\
\nu^2 &=& (1-f)\lambda + f^2 = \lambda^2 -f^2\,.
\end{eqnarray}
A Regge trajectory $J(M^2)$ from (\ref{M_0:simple}) and (\ref{J:simple}) is then linear and its slope $\alpha'$ is
\begin{equation}
\alpha' = {J\over M_0^2} = {\nu\over a(\lambda+1-f)^2} \,. \label{slope}
\end{equation}
The value for $a\alpha'$ depends only on the fraction of scalar confinement. Two important special cases are:

\begin{itemize}
\item[a)] pure scalar confinement $(f=1)$ $\lambda=\sqrt2$, $\nu=1$
\begin{equation}
\alpha' = {1\over 2a} \,.
\end{equation}
In this case we observe from the circular orbit condition (\ref{circ-orbit}) that even a zero mass quark travels at a fraction of the speed of light,
\begin{equation}
v_0 = 1/\sqrt2\,.
\end{equation}
This is because the scalar interaction acts as if it were massive.

\item[b)] pure vector confinement $(f=0)$ $\lambda=\nu=1$
\begin{equation}
\alpha' = {1\over 4a} \,.
\end{equation}
From the circular orbit condition (\ref{circ-orbit}) we see now that as $m\to0$
\begin{equation}
m\gamma_0 \to ar_0 \,,
\end{equation}
so the quark's velocity goes to unity in the $m=0$ limit.
\end{itemize}

For a general value of $f$ we show in Fig.~1 the numerical result for $a\alpha'$ as computed from~(\ref{slope}). We observe that $a\alpha'$ rises nearly linearly over the range $0\le f\le 1$. For $f$ outside this range $\alpha'$ remains monotonic, falling asymptotically to zero for negative $f$ and continuing to rise for $f>1$. A physically appealing model of a heavy-light meson in a highly excited orbital state consists of a straight rotating flux tube with one end fixed. The Regge slope for this tube model is\cite{OV}
\begin{equation}
a\alpha'_{\rm tube} = 1/\pi \,. \label{tubemodelslope}
\end{equation}
From Fig.~1, and the exact relation (\ref{slope}), we see that the slope (\ref{tubemodelslope}) can be mimicked by the choice $f=0.2526\dots$.

We next examine the radial excitation energy implied by our general linear confinement~(\ref{genlinconf}). From the expression (\ref{M^2}) using the result (\ref{omega}) for $\omega$,
\begin{equation}
M_0\omega = (\lambda+1-f)ar_0 {\left(3\nu^2+2f-1\right)^{1/2}\over\lambda r_0}
\end{equation}
and hence
\begin{eqnarray}
\alpha' M^2 &=& J + (2n+1) C \,, \label{genhl}\\
C &=& {\nu \sqrt{3\nu^2+2f-1}\over \lambda(\lambda+1-f)}\,,\\
a\alpha' &=& {\nu\over (\lambda+1-f)^2} \,.
\end{eqnarray}
In Fig.~2 we display the radial energy factor $C$ as a function of $f$. Over the range $0\le f\le 1$, $C$ increases from $1/\sqrt2$ to 1.

As we can see from the general heavy-light results (\ref{genhl}), degeneracy between $J$ and $n$ occurs only for $C={1\over2}, 1, \dots$. From Fig.~2 it is clear that $C=1$ is the only degenerate possibility which can be realized and this occurs when $f=1$ corresponding to pure scalar confinement (for all $f$, $-\infty<f<\infty$, $C$ falls in the range $1/\sqrt2\le C\le\sqrt{3/2}$). The physical consequence of degeneracy is the prediction of ``meson towers", where mesons of different angular momenta lie on top of each other. For scalar confinement $C=1$ and the degeneracy is
\begin{equation}
J+2n = \rm constant \,,
\end{equation}
so for a fixed $M^2$ the next radial excitation will be two units of $J$ different.

\section{Two Light Quark Mesons}

We now proceed to the case of mesons composed of two equal mass quarks\cite{goebel}. The center of momentum point lies equidistant between the quarks which have a relative displacement $r$. The Lagrangian describing the classical motion is then 
\begin{equation}
L = -(2m+S(r)) \sqrt{1-{1\over4}v^2\;} - V(r) \,,
\end{equation}
where the velocity of one quark is given by ${1\over2}v$ and
\begin{equation}
v^2 = \dot r^2 + \omega^2 r^2 \,.
\end{equation}
The canonical momenta are
\begin{eqnarray}
p_r &=& {\partial L\over\partial \dot r} = (2m + S)\gamma {1\over4}\dot r \,,\\
J &=& {\partial L\over\partial\omega} = (2m+S) \gamma{1\over4}\omega r^2\,,
\end{eqnarray}
where $\gamma = \left(1-{1\over4}v^2\right)^{-1/2}$ with the Hamiltonian,
\begin{equation}
H = (2m+S)\gamma + V \,.
\end{equation}
Noting that $p^2=p_r^2+J^2/r^2=(2m+S)^2 {1\over4}(\gamma^2-1)$ we have
\begin{equation}
H = \sqrt{(2m+S)^2+(2p)^2} + V \,. \label{2-part-ham}
\end{equation}

Comparing the above two-particle Hamiltonian (\ref{2-part-ham}) with the single-particle Hamiltonian~(\ref{class-ham}) we note that, in the $m\to0$ limit, the momentum must be half the one-particle momentum to give the same Hamiltonian since $p^2=p_r^2+J^2/r^2$. This means that for a given energy both the radial and the angular momentum are one half the one light quark case. The latter implies that the Regge slope is one half the one-particle slope, or
\begin{equation}
\alpha'(\mbox{two light quarks}) = {1\over2} \alpha'(\mbox{one light quark})\,.
\end{equation}
In addition, the radial momentum is one half the one light quark case which means that the radial frequency $\omega$ is twice the heavy-light frequency. The result is that again the meson spectrum can be written as
\begin{equation}
\alpha'_{\rm LL} M^2 = J + (2n+1)C\,,
\end{equation}
where $C={\nu\sqrt{3\nu^2+2f-1}\over \lambda(\lambda+1-f)}$, exactly the same as in the heavy-light case (\ref{genhl}). In the two light quark meson though,
\begin{equation}
a\alpha'_{\rm LL} = {\nu\over2(\lambda+1-f)^2} = {1\over2} a\alpha'_{\rm HL}\,.
\end{equation}
In this case $M$ is the actual meson mass whereas in the heavy-light case (\ref{genhl}) we must add the heavy quark mass  to $M$.

\section{Comments and Conclusions}

We have demonstrated that both heavy-light and light-light mesons, in dominantly orbital states (DOS), exhibit a universal pattern of excitation. In particular the ratio of radial to orbital excitation energy depends only on the type of linear confinement. The scale of excitation, e.g.\ the Regge slope, also depends on the confinement but is exactly twice as large for the heavy-light mesons as for the light-light mesons. 

In the DOS regime orbital-radial degeneracy, giving rise to meson towers, only appears for scalar confinement. In this case the orbital excitation energy is half the radial energy reminiscent of the non-relativistic harmonic oscillator potential. At low angular momentum the quark's ``penetrating orbits" lower the radial states relative to the orbital states due to the attractive short range interaction.

Although we have used a semi-classical method of description it should give reliable results in the DOS sector. An obvious defect is the neglect of quark spin which even at large range can give a spin-orbit energy. If such a multi-component wave equation possesses leading Regge structure it will be the same as in this work. It turns out that for scalar confinement in the Salpeter equation the leading term does in fact cancel whereas it does not in the Dirac equation\cite{OVW}.

Another possible area of extension is to quark interactions of a four-potential type. The consideration of a spatial vector potential would include flux tube models. For this special case the Regge slope factor of two has already been noted\cite{OV}. It has also been observed that the flux tube exhibits the same $J+2n$ degeneracy as pure scalar confinement\cite{OV}.

\section*{Acknowledgments}
This research was supported in part by the U.S.~Department of Energy under Grant No.~DE-FG02-95ER40896 and in part by the University of Wisconsin Research Committee with funds granted by the Wisconsin Alumni Research Foundation.

\newpage

\begin{figure}
\centering
\hspace{0in}\epsffile{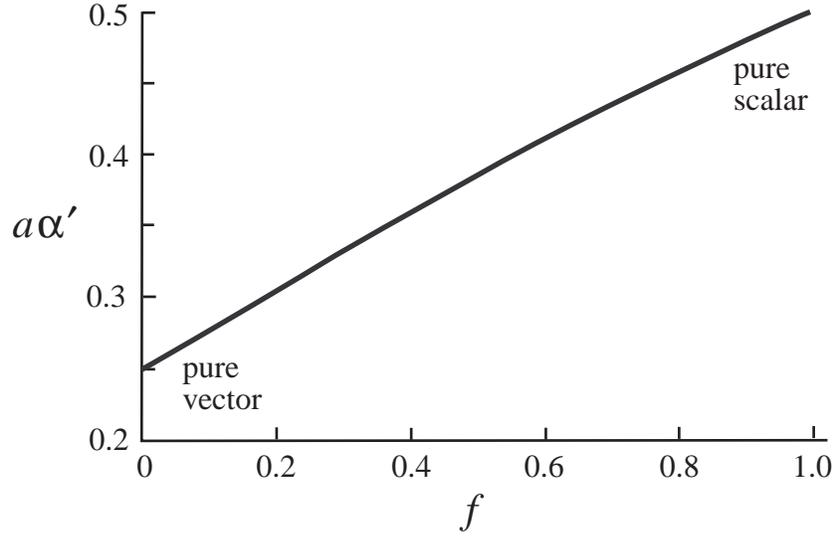}

\caption{Dependance of the Regge slope for the light degrees of freedom of a heavy-light meson on the fraction of scalar confinement $f$. The scalar confinement is taken as $S(r)=far$ and the time-component vector confinement as $V(r)=(1-f)ar$.}
\end{figure}

\newpage
\begin{figure}
\centering
\hspace{0in}\epsffile{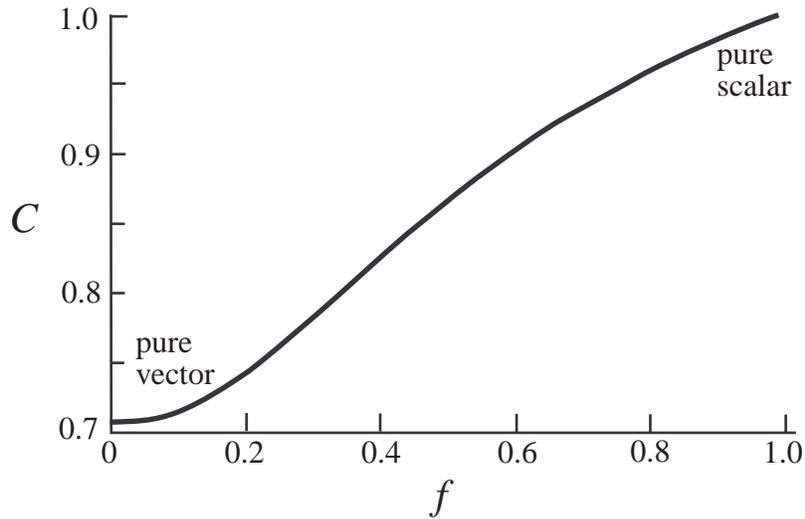}

\caption{Factor $C$ giving the ratio of radial to orbital energies as a function of the fraction of scalar confinement $f$.}

\end{figure}

\end{document}